\documentclass[aps,prd,superscriptaddress]{revtex4}
\usepackage{amssymb}
\usepackage{amsmath}
\usepackage{amsfonts}
\usepackage{epsfig}
\usepackage{graphicx}
\usepackage{epstopdf}
\usepackage{tabularx}
\usepackage{color}
 \usepackage{orcidlink}
\usepackage{amsmath}
\usepackage{slashed}
\usepackage[normalem]{ulem}

\newcommand{\seq}{\begin{subequations}}
\newcommand{\sen}{\end{subequations}}
\newcommand{\eq}{\begin{eqnarray}}
\newcommand{\en}{\end{eqnarray}}

\begin{document}

\title{Nonlocal Portal to the Dark Sector}

\author{Sergey Kovalenko\, \orcidlink{0000-0002-8518-2282}}
\affiliation{Millennium Institute for Subatomic Physics
at the High-Energy Frontier (SAPHIR) of ANID, \\
Fern\'andez Concha 700, Santiago, Chile}
\affiliation{Center for Theoretical and Experimental Particle Physics,
Facultad de Ciencias Exactas, Universidad Andres Bello,
Fernandez Concha 700, Santiago, Chile}

\author{Sergey Kuleshov\,\orcidlink{0000-0002-3065-326X}}
\affiliation{Millennium Institute for Subatomic Physics at
the High-Energy Frontier (SAPHIR) of ANID, \\
Fern\'andez Concha 700, Santiago, Chile}
\affiliation{Center for Theoretical and Experimental Particle Physics,
Facultad de Ciencias Exactas, Universidad Andres Bello,
Fernandez Concha 700, Santiago, Chile}

\author{Valery E. Lyubovitskij\,\orcidlink{0000-0001-7467-572X}}
\affiliation{Institut f\"ur Theoretische Physik, 
Universit\"at T\"ubingen, \\
Kepler Center for Astro and Particle Physics, \\ 
Auf der Morgenstelle 14, D-72076 T\"ubingen, Germany}
\affiliation{Millennium Institute for Subatomic Physics
at the High-Energy Frontier (SAPHIR) of ANID, \\
Fern\'andez Concha 700, Santiago, Chile}

\author{Alexey S. Zhevlakov\,\orcidlink{0000-0002-7775-5917}}
\affiliation{Bogoliubov Laboratory of Theoretical Physics, JINR,
 141980 Dubna, Russia} 
\affiliation{Matrosov Institute for System Dynamics and 
 Control Theory SB RAS, \\  Lermontov street, 134,
 664033, Irkutsk, Russia } 

\begin{abstract}
We propose a nonlocal realization of the Stueckelberg portal
between the Standard Model and Dark Sector, which decouple 
in the local limit. This implies that the mediator, $U(1)_{D}$
Dark Photon $A'$ with a Stueckelberg mass, interacts nonlocally
with the Standard Model quarks and leptons.
We study phenomenological implications of this scenario for
the meson decays into semi-invisible and invisible channels.
We discuss the experimental limitations on the model parameters,
including the nonlocality scale. 

\end{abstract} 

\maketitle
	
\section{Introduction}
\label{sec:introduction}

Light, feebly coupled vector bosons appear ubiquitously in extensions of
the Standard Model (SM), motivated by Dark Sectors (DSs), anomaly--free
$U(1)$ gauge symmetries, and mediators connecting visible and hidden degrees
of freedom. A particularly well-studied benchmark is the ``Dark Photon''
scenario, in which a new Abelian gauge boson interacts with the SM through
kinetic mixing with hypercharge or electromagnetism~\cite{Okun:1982xi,Holdom:1985ag,Essig:2013lka,Fabbrichesi:2020wbt}.
In this work we focus on a complementary and phenomenologically distinct
setup: a massive Dark Photon  $A'_\mu$ as a $U(1)'$ gauge boson
acquiring its mass via the  Stueckelberg mechanism. The latter also leads
to flavor--nondiagonal $A'-$SM fermion couplings---Stueckelberg portal (StP) 
introduced in Ref.~\cite{Kachanovich:2021eqa}. 
The  kinetic mixing, if introduced, can be absorbed by the flavor--diagonal
vector couplings by an appropriate field redefinition. Moreover, we assume
that these $A'-$SM interactions are nonlocal with a characteristic
nonlocality scale $\Lambda_{\rm NL}$ related to some UV, possibly
super-Planckian, physics. As revealed in Ref.~\cite{Krasnikov:1987mz},  
the nonlocality allows, in the case of the kinetic mixing, reconciling  
the tension between the observable DM relic density and nonobservations
DM signal in direct--detection experiments. 

Depending on the spectrum of possible hidden--sector states, $A'$ can decay
visibly into lepton flavor--conserving as well as flavor--violating channels
$A'\to \ell_1^{\pm} \ell_2^{\mp}$  with $\ell=e,\mu,\tau $ and into hadrons
or dominantly invisibly $A'\to \chi\bar{\chi}$ into the DS 
particles. The hadronic phenomenology is controlled by the embedding of
the quark currents in into low-energy QCD degrees of freedom. This makes
rare meson processes one of the primary probes of the model for $m_{A'}$
ranging from the sub-MeV up to the multi-GeV scale, and yields a set of
constraints and signatures that can be different from the usual kinetically
mixed Dark Photon scenario.

Our goal is twofold. First, we match the quark-level $A'$ nonlocal
interactions onto an effective theory of pseudoscalar and vector mesons
including the anomalous Wess--Zumino--Witten (WZW) terms.
Second, using this effective framework, we compute a set of benchmark amplitudes
and decay widths relevant to laboratory searches, including radiative
semi-invisible pseudoscalar meson decays and invisible vector--meson decays
with missing energy lost in $A'$ decays into the DS particles.
These observables allow for a consistent recasting of existing limits and
provide a map between fermion-level couplings and accessible meson
phenomenology. We analyze the impact of the nonlocality of $A'-$SM
interactions and estimate the corresponding scale of  nonlocality.

The paper is organized as follows. In Sec.~\ref{sec:setup}, we formulate the
quark--level nonlocal $A'$ model with the assumptions and conventions we adopt.
In Sec.~\ref{sec:chiral_matching}, we match the quark currents onto the chiral
and resonance (vector-meson) effective theory, including the WZW--induced
interactions relevant for anomalous transitions.
In Sec.~\ref{sec:Mes-Phen-DPh}, we derive decay widths and branching ratios
for the key mesonic channels and discuss characteristic
kinematic regimes (on-shell versus off-shell $A'$ production). 
Also we assemble and interpret the corresponding constraints, distinguishing
visible and invisible $A'$ decay scenarios, and summarize our analysis. 
In  Sec.~\ref{sec:Discussion}, we briefly discuss existing constraints
on Dark Photons from the perspective of our nonlocal model. 
Finally, in Sec.~\ref{sec:Conclusions}, we present the summary of our
results and findings. 

\section{Theoretical setup}
\label{sec:setup}

We consider a setup involving two isolated sectors---the SM 
sector and the DS mutually singlet or sterile with respect to gauge symmetries
of each other. For the DS, we assume a simplified structure based
on the $U(1)_D$ gauge group with the scalars $S$ and vectorlike fermions
$\chi$ charged under $U(1)_D$ and sterile with respect to the SM gauge
groups. In this scenario, all the SM fields have $U(1)_D$ charges zero.
The model gauge group is 
\eq 
G \equiv SU(3)_C \times SU(2)_L \times U(1)_Y \times U(1)_D 
\en
The relevant part of the model Lagrangian is
\eq 
 \label{eq:Lagr-1}
\mathcal{L} &\supset&
 -\frac{1}{4} B^{\mu\nu} B_{\mu\nu} -\frac{1}{4} {X}^{\mu\nu} {X}_{\mu\nu} 
 -\frac{1}{2} B_{\mu\nu} {\epsilon}_Y {X}^{\mu\nu}
 - \frac{1}{2} M^2_{X} X^\mu X_\mu \\
 \nonumber
 &+&\bar\psi\, ( i\slashed{\partial} - m_\psi) \, \psi + 
 \bar\chi\, ( i\slashed{\partial} - m_\chi) \chi +  e' X^\mu \, \bar\chi\,
 \gamma_\mu  \chi 
\en
where  
\eq 
{X}_{\mu \nu} \equiv \partial_{\mu} {X}_{\nu}-\partial_{\nu} {X}_{\mu}, \ \ \
{B}_{\mu \nu} \equiv \partial_{\mu} {B}_{\nu}-\partial_{\nu} {B}_{\mu}
\en
where $B_\mu$ and $X_\mu$ are the gauge fields of the Abelian groups
$U(1)_Y$ and $U(1)_D$, respectively.
Note that without introduction of the  SM$-$DS portal, the SM and DS are
completely decoupled from each other even at the quantum level.
Neither SM nor DS interactions can generate the cross terms.
The gauge kinetic terms in Eq.~(\ref{eq:Lagr-1})
can be diagonalized by rotating the gauge fields (for recent review
see, e.g., Ref.~\cite{Fabbrichesi:2020wbt}). At the limit $\epsilon_Y\ll 1$
we have  
\eq 
{B}_{\mu} = A_{\mu}+\epsilon A_{\mu}^{\prime}, \ \ \ 
{X}_{\mu} = A_{\mu}^{\prime}-\epsilon \tan \theta_{W} Z_{\mu} 
\en
with $\epsilon= \epsilon_Y \cos\theta_{W}$. Here,
$A_\mu$ and $Z_\mu$ are the usual neutral electroweak gauge
bosons---photon and $Z^0$-boson, respectively.  
Then, in the $A'$ sector we have 
\eq 
    \label{eq:Lag-SSB-1}  
    \mathcal{L} \supset &-&  e \, \hat\epsilon \, A_{\mu}^{\prime} 
    \sum_{i} Q_{\psi_i} \bar\psi_i \gamma^\mu \psi_i 
    -g_D A_{\mu}^{\prime} \bar{\chi} \gamma^{\mu} \chi 
    + \frac{1}{2} M_{A^{\prime}}^{2} A_{\mu}^{\prime}A^{\prime \mu}
    + A'_\mu \, \sum\limits_{ij} \, \bar\psi_i \, 
\gamma^\mu \, \left(g^{v}_{ij} + g^{a}_{ij} \gamma_5  \right) \, \psi_j\,, 
\en
where $\psi_i = q, \ell$ are the SM quarks and leptons with electric
charges $Q_{\psi_i}$ and $M_{A'}=M_{X}$. Here, the last dimension-4 operator,
with $g^{v,a}_{ij}$ flavor--nondiagonal couplings, originates from
a dimension-5 operator arising in the Stueckelberg realization of $U(1)_D$
symmetry group. These flavor--nondiagonal SM$-$DS portal was recently found
in Ref.~\cite{Kachanovich:2021eqa}. Once this StP portal 
is included the first term, originating from the kinetic $A-A'$ mixing,
can be eliminated by  the redefinition of the flavor-diagonal vector
coupling $g^v_{ii} \to g^v_{ii}+e\epsilon$.  

Here, we will focus on the StP as a unique communication between
the SM and DS. Note that the $A-A'$ kinetic mixing is regenerated by
StP interactions in~\eqref{eq:Lag-SSB-1} via fermion loop corrections.
This mixing can be eliminated through appropriate $A$ and $A'$ field
redefinition and absorption into couplings $g^{v,a}$. This results their
renormalization scale $\mu$ dependence controlled by the well-known
renormalization-group equation running. We choose the scale $\mu=\mu_0$
at which the kinetic mixing is vanishing. In the phenomenological treatment
we can always restore the conventional case of the pure kinetic mixing
identifying $e\epsilon = g^{(v)}$ and imposing $g^{(v)}_{ii}= g^{(v)}$,
$g^{(v)}_{i\neq j} =0$, $g^{(a)}_{ij}=0$ for $\forall  \, i,j$. 

In this setup, we will explore a scenario assuming that this communication
is \textit{nonlocal} while the SM and the DS  separately are described by
local theories. This is an extension of the idea about nonlocality of
the kinetic mixing, recently proposed in Ref.~\cite{Krasnikov:2024fdu}.
The nonlocality is implemented by promoting the couplings $g^{v,a}$
in Eq.~(\ref{eq:Lag-SSB-1}) to a differential operators.
We assume the universal nonlocality replacing 
\eq 
\label{eq:NL-portal-1}   
g^{v,a}_{ij}\to  \hat\epsilon(\partial^2) \, g^{v,a}_{ij} \, ,
\en
which, in the matrix elements of physical processes, gives rise
to a \textit{vertex form factor}
$\hat\epsilon(k^2)$ depending on the squared momentum transfer $k^2$. 

\begin{figure}[t]
	\centering
        \includegraphics[width=0.6\linewidth]{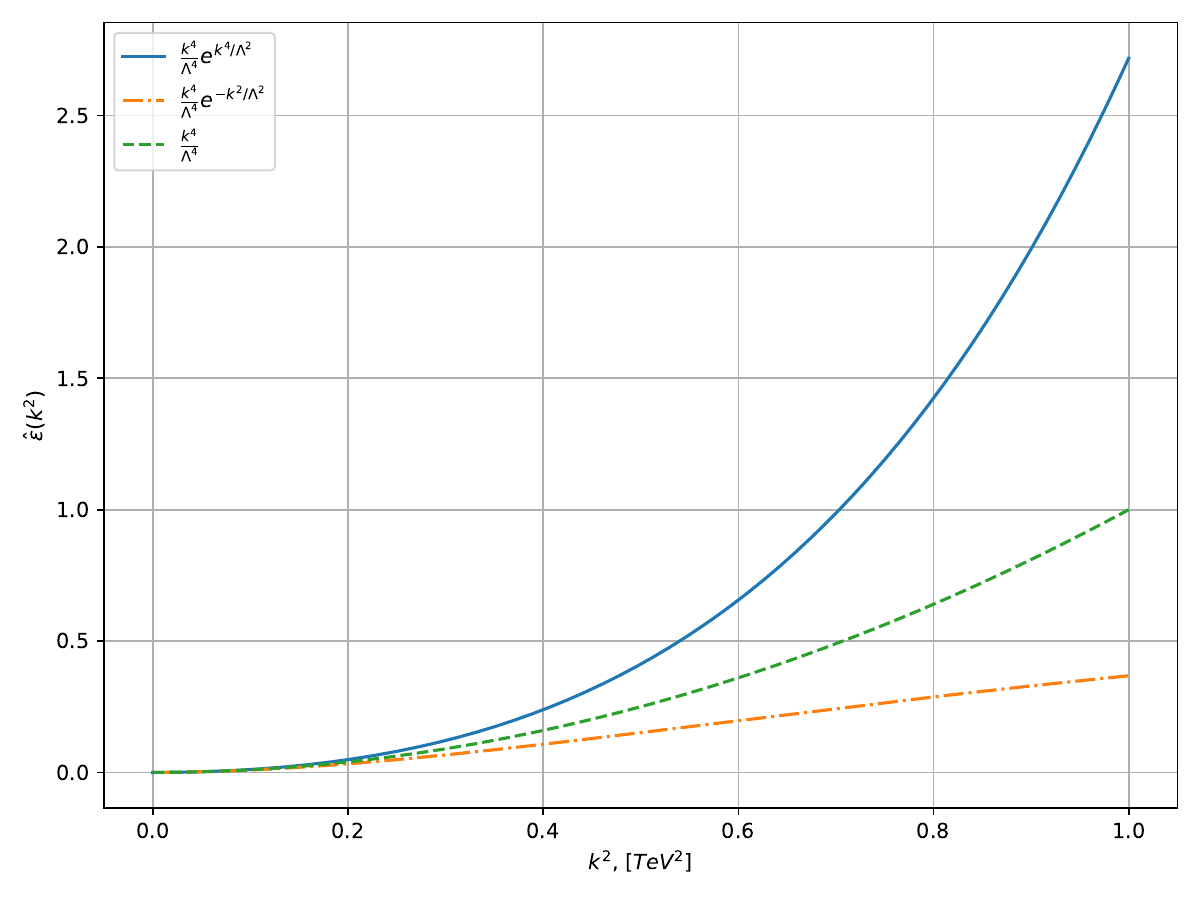}
	\caption{Three examples of the nonlocal form factor 
        $\hat\epsilon(k^2)$
        at the nonlocality scale $\Lambda_{\rm NL}=1$ TeV.} 
	\label{fig:eps_k2}
\end{figure}

For a nonlocal theory to be physically tractable, this form factor
must satisfy certain conditions~\cite{Efimov:1967pjn,Efimov:1993zg}, 
in particular, it must be an entire function (for more recent publications
see Refs.~\cite{Krasnikov:2024fdu,Krasnikov:2022kgl,Biswas:2014yia}). 
We choose it in the  following simplified form 
\eq 
\label{eq:NL-portal-2}
 \hat\epsilon(k^2) = 
\left(\frac{k^2}{\Lambda_{\rm NL}^2} \right)^2 \,
 \exp\left(\frac{k^2}{\Lambda_{\rm NL}^2}\right)\,. 
\en
Here, $\Lambda_{\rm NL}$ is a scale referring
to a fundamental physics behind nonlocality of the SM-DS communication.
Also, in the limit $\Lambda_{\rm NL} \to \infty$, the nonlocal
portal specified by Eqs.~(\ref{eq:NL-portal-1}) and (\ref{eq:NL-portal-2})
is turning off. In Euclidean momentum space, this form factor provides
ultraviolet and infrared convergence of the loop diagrams.
Here, the ultraviolet and infrared convergences
imply the matrix elements to be finite at large loop momentum
and vanishing masses of the SM fermion, respectively.
In Fig.~\ref{fig:eps_k2}, we show the $k^2$ behavior of the nonlocal
form factor for the three different
analytic forms and at the nonlocality scale $\Lambda_{\rm NL}=1$ TeV.

Note that this type of nonlocality cannot be generated by local renormalizable
interactions mediated by particle exchange. Note, nonlocality via an infinite
number of local fields was proposed in Ref.~\cite{Krasnikov:1987mz}.
Its fundamental origins should reside among nonlocal objects like strings,
topological defects, or compact manifolds in extra-dimensional theories.
In this case, one expects that $\Lambda_{\rm NL}$ is of the order of
the Planck scale $\Lambda_{\rm Pl}$. Importantly, some of these frameworks
do not exclude sub-Planckian values for the nonlocality scale
$\Lambda_{\rm NL}$, which, in certain scenarios such as large extra dimensions,
may lie in the TeV ballpark.

\section{Dark Photon Interactions with mesons}
\label{sec:chiral_matching}

In order to study vector meson $V = \rho,\omega,\phi$ and the light
pseudoscalar $P=\pi^0,\eta, \eta'$ meson decays as well as their
contribution to the processes with Dark Photon, we use the meson effective
field theory framework. To this end we integrate out the light quark
fields $u$, $d$, and $s$ in the Lagrangian~(\ref{eq:Lag-SSB-1}) embedding it
to the low-energy effective Lagrangian based on ideas of chiral
perturbation theory (ChPT)~\cite{ChPT}. The terms relevant for the $A'-$light
meson phenomenology include the $A'-V$ kinetic mixing (or, equivalently,
mass mixing), the WZW plus vector--meson dominance.
Explicitly we have
\eq 
\label{eq:Lag-eff-V-P-1}  
{\cal L} &\supset&   \frac{1}{2}  \, g_{VA'}  \, V_{\mu\nu}\,
\hat\epsilon F^{\prime \mu\nu} + 
\frac{e^2}{4}  g_{P\gamma A'} \,  
\varepsilon_{\mu\nu\alpha\beta} \, P \,  
F^{\mu\nu} \, \hat\epsilon F^{\prime\, \alpha\beta} +
\frac{e}{4}  \, g_{VP A'} \,  
\varepsilon_{\mu\nu\alpha\beta} \, P \,  
V^{\mu\nu} \, \hat\epsilon F^{\prime\, \alpha\beta} + \mbox{ H.c.} \,,
\en
where $V_{\mu\nu} = \partial_\mu V_\nu -\partial_\nu V_\mu$ is the stress
tensor of vector meson field $V_\mu$. In what follows we are studying
a simplified case of the $A'qq$ vector couplings $g_V = g_{A'} Q_q$,
where $Q_q$ is the quark electric charge
and $g_{A'}$ is a universal parameter for the light quarks.
For this benchmark case we have 
\eq 
\label{eq:eff-couplings-1}
    g_{VA'}= g_{V\gamma} g_{A'}\, \quad  
    g_{P\gamma A'} = g_{P\gamma\gamma} g_{A'}\,, \quad 
    g_{VP A'} = g_{VP\gamma} g_{A'} \,.
\en
The effective couplings $g_{V\gamma}$, $g_{P\gamma\gamma}$, and $g_{VP\gamma}$
can be extracted from data on the corresponding meson decays and
the decay rate formulas
\eq 
\label{eq:V-GG-123}  
\Gamma(V \rightarrow e^+ e^-) &=& \frac{4 \pi}{3} \,
\alpha^2 \, g_{V\gamma}^2\,  M_V \,, \\
\label{eq:P-GG-DecRate-1}
    \Gamma(P \rightarrow \gamma\gamma)
&=& \frac{\pi \, \alpha^2}{4} \, g^2_{P\gamma\gamma}
\, m_P^3 \\
\Gamma(V \rightarrow P\gamma)
&=& \frac{\alpha}{24} \, g^2_{VP\gamma}
\, m_V^3 \, (1 - x_{PV})^3 \\
\Gamma(P \rightarrow V\gamma)
&=& \frac{\alpha}{8} \, g^2_{VP\gamma}
\, m_P^3 \, (1 - x_{VP})^3 \,,
\en
where $M_{V}$ and $M_P$ are the masses of the vector $V=\rho^0,\omega,\phi$
and pseudoscalar $P=\pi^0,\eta, \eta'$ mesons, respectively. 
Here and in the following we denote by $x_{ij} = m^2_i/m_j^2$ the ratio
of squares of masses $m_i$ and $m_j$. 

The values of the couplings $g_{V\gamma}$, $g_{P\gamma\gamma}$,
and $g_{VP\gamma}$
can be found in Ref.~\cite{ParticleDataGroup:2024cfk}.
In our analysis, we use their central values:  
\begin{align}
\label{gP2g}
g_{\rho^0\gamma}     &= 0.202\,,
& g_{\omega\gamma} &= 0.059\,,
& g_{\phi\gamma}   &= 0.075\,,
\nonumber\\
g_{\pi^0\gamma\gamma} &= g_{\eta\gamma\gamma} = 0.274 \, {\rm GeV}^{-1}\,,
& g_{\eta'\gamma\gamma} &= 0.341 \, {\rm GeV}^{-1}\,,
&
\nonumber\\
g_{\rho^0\pi^0\gamma}    &= 0.699 \, {\rm GeV}^{-1}\,, 
& g_{\rho^0\eta\gamma}  &= 1.55 \, {\rm GeV}^{-1}\,, 
& g_{\rho^0\eta\gamma}  &= 2.73 \, {\rm GeV}^{-1}\,,
\nonumber\\
g_{\omega\pi^0\gamma}    &= 2.33 \, {\rm GeV}^{-1}\,, 
& g_{\omega\eta\gamma}  &= 0.45 \, {\rm GeV}^{-1}\,,  
& g_{\omega\eta'\gamma} &= 0.45 \, {\rm GeV}^{-1}\,,
\nonumber\\
g_{\phi\pi^0\gamma}     &= 0.136 \, {\rm GeV}^{-1}\,,   
& g_{\phi\eta\gamma}    &= 0.67 \, {\rm GeV}^{-1}\,,    
& g_{\phi\eta'\gamma}   &= 0.706 \, {\rm GeV}^{-1}\,.
\end{align}

\section{Meson Phenomenology of Dark Photon}
\label{sec:Mes-Phen-DPh}

Here, we analyze some phenomenological aspects of Dark Photon in the sector
of light mesons and compute amplitudes and decay widths of benchmark
processes on the basis of the effective
hadronic Lagrangian~\eqref{eq:Lag-eff-V-P-1}. 

\subsection{$A'$ decay to lepton-antilepton pair}

\begin{figure}[ht!]
\begin{center}
\includegraphics[trim={1cm 23.5cm 1.75cm 1cm}, clip, scale=0.65]{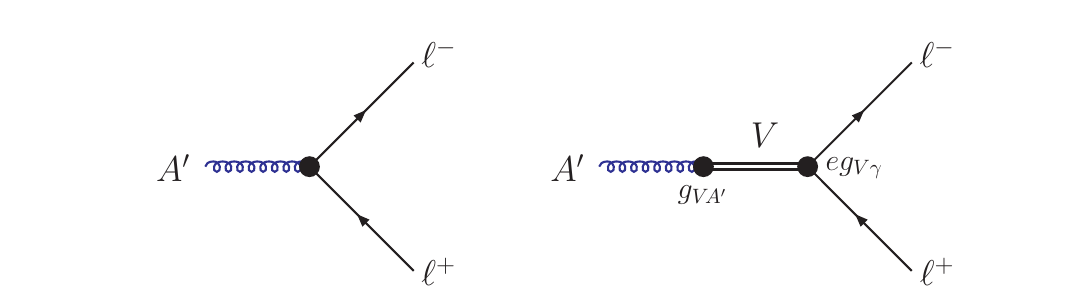}
	          \caption{Feynman diagrams contributing to
                    $A'$ decay to lepton-antilepton pair:
                    direct and resonance diagram. 
                  \label{figA2ell}}
                  \end{center}
                  \end{figure}

We start with the $A'$ decay to lepton-antilepton pair.
There are two diagrams contributing to this decay: direct (left panel)
and resonant via the Dark Photon--vector meson mixing
(right panel) diagrams, shown in Fig.~\ref{figA2ell}.

The corresponding matrix element ${\cal M}(A' \rightarrow \ell^+\ell^-)$
is given by the sum of direct ($D$) and resonance ($R$) contributions:
\eq
   {\cal M}(A' \rightarrow \ell^+\ell^-) =
   {\cal M}_{\rm D}(A' \rightarrow \ell^+\ell^-)
 + {\cal M}_{\rm R}(A' \rightarrow \ell^+\ell^-) \,. 
 \en
 which we calculate from the Lagrangian~(\ref{eq:Lag-eff-V-P-1}).
 For the direct contribution, we have
\eq 
\label{eq:MinvAprepem} 
{\cal M}_{\rm D}(A' \rightarrow \ell^+\ell^-) &=&  \varepsilon_\mu(p) \,
 \, \bar u(q_1) \, \gamma^\mu \, 
\left(G^{v}_{\ell\bar\ell} + G^{a}_{\ell\bar\ell} \gamma^5\right) 
\, v(q_2) 
\en
where $\bar u(q_1)$ and $v(q_2)$ are spinors of lepton with momentum
$q_1$ and antilepton with momentum $q_2$, respectively;
$\varepsilon_\mu(p)$ is polarization vector of the Dark Photon.  
Here, we denoted the nonlocal vector and axial
couplings of $A'$ with leptons as 
$G^{v,a}_{\ell\bar\ell} = \epsilon(m_{A'}^2) \,  g^{v,a}_{\ell\bar\ell} \,.$

The contribution of the resonance diagram is
\eq 
{\cal M}_{\rm R}(A' \rightarrow \ell^+\ell^-) &=& 
e^2 \, \epsilon_\mu(p) \, g_{A'\ell\bar\ell}^R \, 
\bar u(q_1) \, \gamma^\mu \, v(q_2) \,, 
\nonumber\\
g_{A'\ell\bar\ell}^R &=&
\sum_{V = \rho^0, \omega, \phi}
\, g_{V\gamma} 
\, \frac{m_{A'}^2 \, g_{VA'}(m_{A'}^2)}{m_V^2 - m_{A'}^2 - i \Gamma_V m_V}
\,,
\en
In order to avoid the $m_V^2\simeq m_{A\prime}^2$ singularity,
we retain the finite width $\Gamma_V$ in the Breit--Wigner propagator.
For convenience, we define $g_{VA\prime}\equiv g_{V\gamma}\,g^v$
as in Eq.~(\ref{eq:eff-couplings-1}).
Summing up the direct and the resonant contributions of
the $V = \rho^0, \omega, \phi$
we get the $A' \rightarrow \ell^+\ell^-$ decay width 
\eq 
\label{eq:A-LAL-1}
\Gamma(A' \rightarrow \ell^+ \ell^-) = \frac{\alpha m_{A'}}{3}
\,
\sqrt{1 - 4 x_{\ell A'}} \, (1 + 2 x_{\ell A'})
\,   \biggl[\left(G^{v}_{\ell\bar\ell}\right)^2
  + \left(G^{a}_{\ell\bar\ell}\right)^2
  + 2 G^{v}_{\ell\bar\ell} Re(g_{A'\ell\bar\ell}^R)
  + |g_{A'\ell\bar\ell}^R|^2
  \biggr] 
\,.
\en
The similar analysis can be straightforwardly extended to the
$e^+ e^-$ collisions and the bremsstrahlung processes with Dark Photon
production. 
These processes are important to search for Dark Matter signals
by the missing energy technique~\cite{NA64:2024nwj,BaBar:2009gco}.  

\subsection{$A'$ production in rare decays of $\pi^0$, $\eta$, and
  $\eta'$ mesons}

\begin{figure}[ht!]
\begin{center}
\includegraphics[trim={1cm 21cm 1cm 3cm}, clip, scale=0.65]{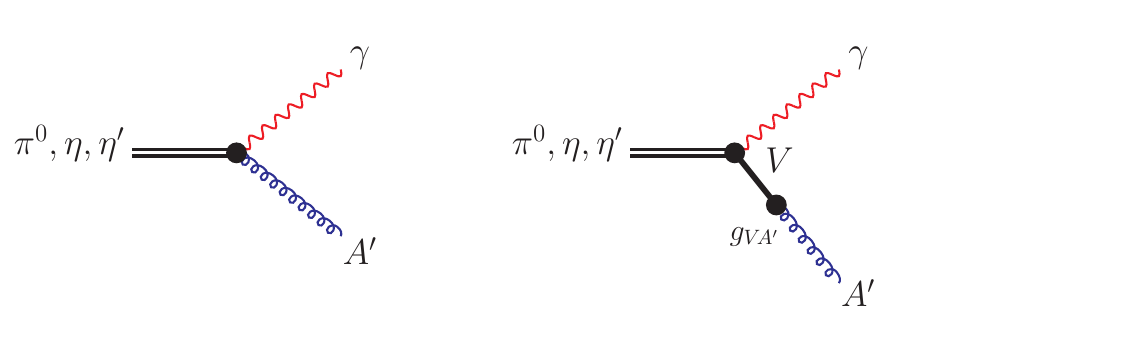}
          \caption{Feynman diagrams contributing to semi-invisible modes 
          of light neutral pseudoscalars $P = \pi^0, \eta, \eta'$.   
          \label{PAprimegamma}}
\end{center}
\end{figure}

Next, we consider the $A'$ production in rare decays of light
pseudoscalar mesons $P=\pi^0,\eta, \eta'$.
In our approach, there are two diagrams
in Fig.~\ref{PAprimegamma} which contribute to this process
at leading order.   
The left diagram describes  the 
direct $A'$ production, while the right one the
resonant, induced by the $V-A'$ transition. 

The total matrix element of the $P \rightarrow \gamma A'$ decay is given by 
\eq
   {\cal M}(P \rightarrow \gamma A')
   = {\cal M}_{\rm D}(P \rightarrow \gamma A')
   + {\cal M}_{\rm R}(P \rightarrow \gamma A') \,,   
\en
where ${\cal M}_{D}$ and ${\cal M}_{R}$ correspond to
direct ($D$) and $V$-resonance ($R$) diagrams in Fig.~\ref{PAprimegamma}, 
respectively. Matrix element is induced by the $PFF'$ anomaly term
in the effective Lagrangian~(\ref{eq:Lag-eff-V-P-1}).
For the local $FF'$-portal scenario, 
$\hat\epsilon \equiv \epsilon$ is a constant to be determined from
the data analysis. In the nonlocal case for the on-shell $A'$, we have
$\hat\epsilon = \epsilon(m_{A'}^2)$. 
For the direct matrix element ${\cal M}_{D}$ of $A'$ production, we have
\eq 
{\cal M}_D &=&
e^2 \, \epsilon_{\mu\nu\alpha\beta} \, \epsilon_\gamma^\mu(q_1) \,
\epsilon_{A'}^\nu(q_2) \, q_1^{\alpha} \, q_2^{\beta} \, g_{P\gamma A'} 
\en
where we denoted $g_{P\gamma A'} = \epsilon(m_{A'}^2) g_{P\gamma A}$.
To calculate resonance matrix element ${\cal M}_R$, we
need to specify interaction Lagrangians describing the couplings $g_{VP\gamma}$ 
of $P = \pi^0$, $\eta$, and $\eta'$ mesons with photon and vector mesons
$V = \rho^0$, $\omega$, and $\phi$.
These couplings occur in the effective Lagrangians given by
\eq
{\cal L}_{VP\gamma} =
\frac{e}{4} \, g_{VP\gamma} \,  \epsilon_{\mu\nu\alpha\beta} \, P \,  
V^{\mu\nu} \, F^{\alpha\beta} \,.
\en  
The $V$-resonance contribution
${\cal M}_{R}$ to the matrix element of the
$A'$ production is given by 
\eq 
{\cal M}_{\rm R} &=&
e^2 \, \epsilon_{\mu\nu\alpha\beta} \, \epsilon_\gamma^\mu(q_1) \,
\epsilon_{A'}^\nu(q_2) \, \, q_1^{\alpha} \, q_2^{\beta} \, g_{P\gamma A'}^R \,,
\nonumber\\
g_{P\gamma A'}^R &=&
\sum_{V = \rho^0, \omega, \phi}
\, g_{VP\gamma} \,
\frac{m_{A'}^2}{m_V^2}  
\, \frac{m_{A'}^2 \, g_{A'V}^2(m_V^2)}{m_V^2 - m_{A'}^2 - i \Gamma_V m_V}
\,. 
\en
The width of the $P  \rightarrow  \gamma A'$ decay is calculated using
the formula 
\eq
\Gamma(P \rightarrow \gamma A')
= \frac{\pi \alpha^2}{4} \, m_P^3 \,
\biggl[g^2_{P\gamma A'}
  + 2 g_{P\gamma A'} Re(g_{P\gamma A'}^R)
  + |g_{P\gamma A'}^R|^2
  \biggr] \,. 
\en 
Note that decays of neutral light pseudoscalar mesons ($\pi^0$, $\eta$,
and $\eta'$) to $\gamma A'$ are under study in the NA64, NA62, and COHERENT
experiments~\cite{NA64:2024klw}-\cite{COHERENT:2022pli}. 

\subsection{Invisible Decays of Vector Mesons}

\begin{figure}[ht!]
\begin{center}
\includegraphics[trim={1cm 23cm 1cm 1cm}, clip, scale=0.65]{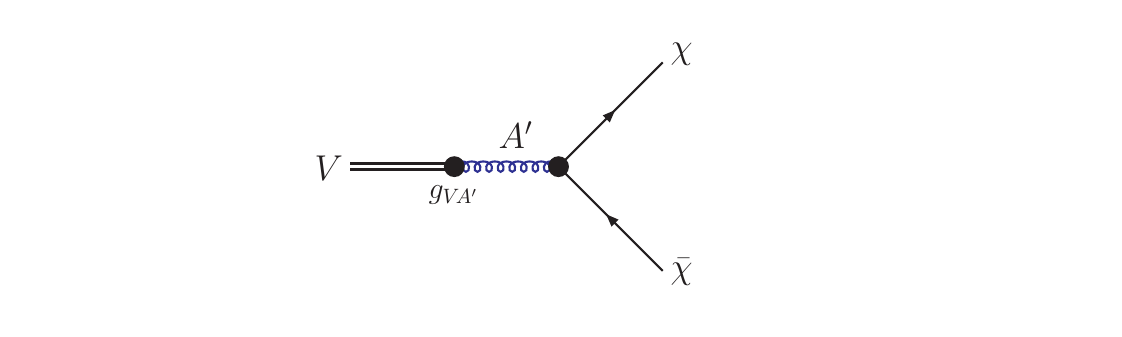}
          \caption{Feynman diagram contributing to invisible modes 
          of neutral vector mesons. 
          \label{VA2chi}}
\end{center}
\end{figure}

The vector mesons  can decay invisibly $V \rightarrow \chi\bar\chi$ to the
Dark fermions $\chi\chi$ due to the $V-A'$ mixing induced by the first term
in \eqref{eq:Lag-eff-V-P-1}. The Feynman diagram contributing to this invisible
decay mode of neutral vector mesons is shown in Fig.~\ref{VA2chi}. 
The second vertex is described by the $g_D A'\chi\chi$ term 
in~(\ref{eq:Lag-SSB-1}). 

Then, the vector--meson decay rates to this invisible mode
is given by 
\eq
\Gamma(V \to \chi\bar\chi)  =
\frac{\alpha_D}{6} \, g_{A' V}^2(m_V^2) 
\, m_V \, \frac{(1 + 2 x_{\chi V}) \sqrt{1 - 4 x_{\chi V}}}
{(1-x_{A'V})^2 + z^2} \,,
\en
where $z = \Gamma_{A'\to \chi\bar\chi}/m_V$,  
$m_V$, $m_{A'}$, and $m_\chi$ are the masses of vector meson,
intermediate Dark Photon
and DM fermion, respectively.  
Here, we use the Breit-Wigner propagator for $A'$ to avoid
$m_{A'}^2 \simeq m_{V}^2$ singularity by assuming that
$A'$ its total width is dominated by the
$A' \to \chi\bar\chi$ invisible mode. 

\subsection{Semi-Invisible Decays of Light Pseudoscalar Mesons}

\begin{figure}[ht!]
\begin{center}
\includegraphics[trim={1cm 20cm -1cm 1cm}, clip, scale=0.65]{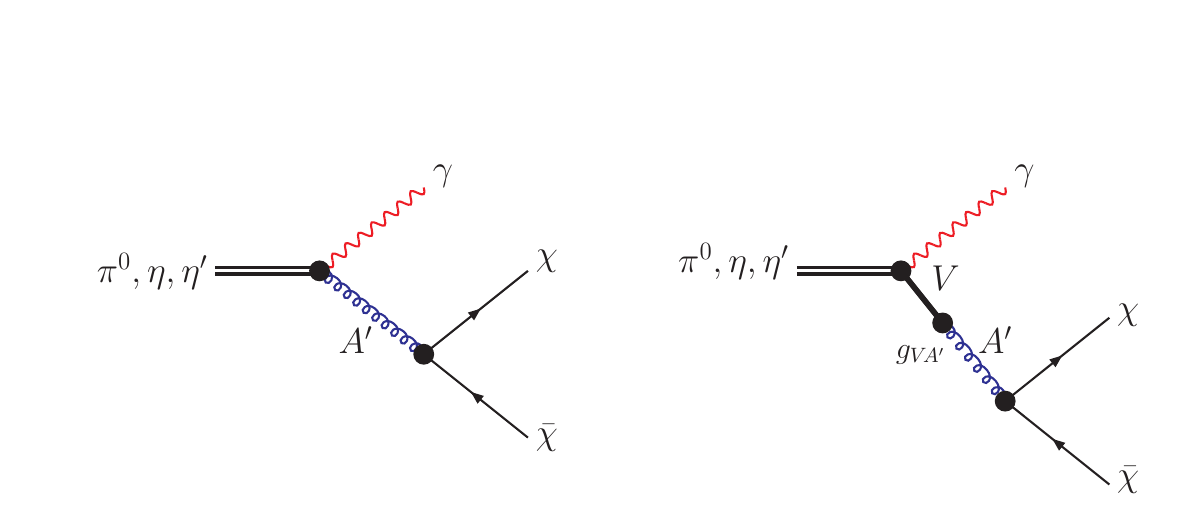}
	         \caption{Feynman diagrams contributing
                 to semi-invisible decays of
                 light pseudoscalar mesons. 
                 \label{PVgchi2}}
\end{center}
\end{figure}

In this section we discuss results for
the semi-invisible decays of light pseudoscalar
mesons $P = \pi^0, \eta, \eta'$. 
Diagrams contributing to these processes are shown
in Fig.~\ref{PVgchi2}. 
Here, as before, we consider two mechanisms:
(i) direct $A'$ production (left diagram) and
(ii) resonant via $V-A'$ transition (right diagram).

Summing up both contributions we obtain
\eq
\label{WithRhoDecayPS}
d\Gamma(P \to \chi\bar\chi\gamma) &=& \frac{\alpha \alpha_D}{24\pi m_P^3}
\frac{(q^2-m_{\eta'}^2)^3 (q^2+2m_\chi^2)(q^2-4m_\chi^2)^{\frac{1}{2}}}
     {(m^2_{A'}-q^2)^2+\Gamma^2_{A'\to \chi\bar\chi}m^2_{A'}} 
\, \frac{dq^2}{\sqrt{q^2}} \nonumber\\
&\times&\biggl[g_{P\gamma A'}^2
+ 2 g_{P\gamma A'} Re(g_{P\gamma A'}^R(q^2))
+ |g_{P\gamma A'}^R(q^2)|^2
\biggr] \,, 
\en
where 
\eq
g_{P\gamma A'}^R(q^2) = 
\sum_{V = \rho^0, \omega, \phi}  
\frac{q^2 \, g_{VP\gamma} \, g_{VA'}(q^2)}{q^2 - m^2_{V}
+ i \Gamma_{V} m_{V}} \,.  
\en 
This is the nonlocal generalization of the result derived
in Ref.~\cite{Gninenko:2023rbf}.

\section{Discussion}
\label{sec:Discussion}

\begin{figure}[ht!]
	\centering
	\includegraphics[width=0.7\linewidth]{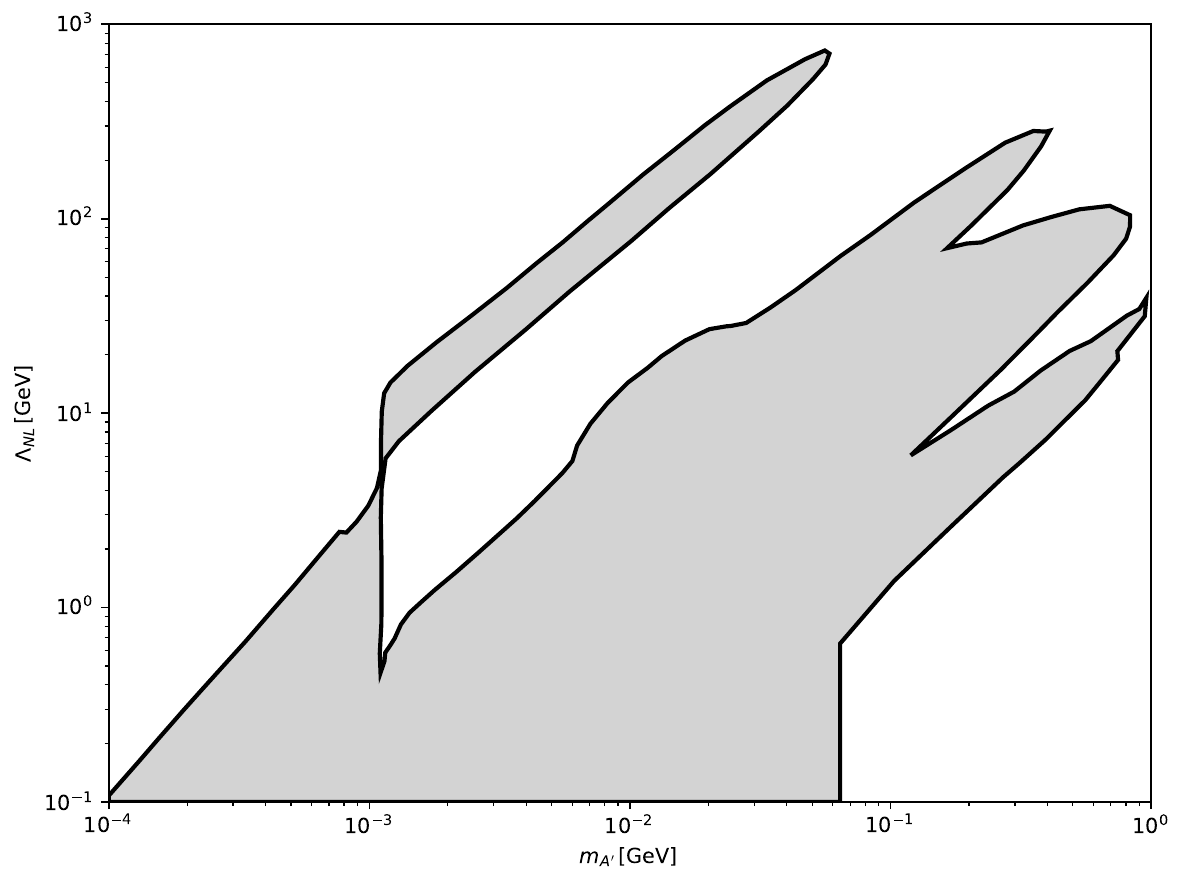}
	\caption{Exclusion plot in the $(m_{A'}, \Lambda_{\rm NL})$ plane.
        for $\epsilon = 1$ derived from Ref.~\cite{Caputo:2025avc}.
        Shaded region is excluded by the observations and calculation from cosmology.} 
	\label{fig:boundslambda}
\end{figure}

Here we briefly discuss the impact of existing experimental bounds in the case 
of the nonlocal realization of the Stueckelberg portal between the SM and DS.  
The main constraints on the Dark Photon parameters are obtained from
$s$-channel processes, where a massive Dark Photon is produced on shell,
corresponding to timelike momentum transfer $k^2 = m_{A'}^2$. 
In the nonlocal portal models, the constant kinetic mixing parameter
$\epsilon_Y$ in~(\ref{eq:Lagr-1}) or, as in the present case,
the Stueckelberg couplings $g^{v,a}_{ij}$ in~(\ref{eq:Lag-SSB-1})
are replaced by momentum-dependent quantities according
to Eq.~(\ref{eq:NL-portal-1}).
Therefore, the bounds existing in the literature on local Dark Photon models,
typically presented in the form $\epsilon < \epsilon_{\rm exp}(m_{A'})$,
can be recast into the constraint
$\hat{\epsilon}(m_{A'}^2;\Lambda_{\rm NL}) g/e < \epsilon_{\rm exp}(m_{A'})$
for processes involving 
$s$-channel photon exchange $k^2 = m^2_{A'}$, resulting in a lower bound curve
in the $(m_{A'}, \Lambda_{\rm NL})$ plane. We show the corresponding exclusion plot
in Fig.~\ref{fig:boundslambda}, derived by applying this approach
to the collider and cosmological bounds on the local Dark Photon collected
in Ref.~\cite{Caputo:2025avc}. 

As seen from Fig.~\ref{fig:boundslambda}, the scale $\Lambda_{\rm NL}$ of 
nonlocal interaction between Dark Photons of mass $m_{A'}$ and SM fermions
weakens in the small $m_{A'}$ limit, although the couplings may remain
significant. This behavior originates from the infrared suppression of
the nonlocal form factor $\hat{\epsilon}(k^2)\propto (k^2/\Lambda_{\rm NL}^2)^2$
in~(\ref{eq:NL-portal-2}), significantly reducing the on-shell production
rate in the regime $m_{A'} \ll \Lambda_{\rm NL}$. Consequently, searches
relying on resonant Dark Photon production lose sensitivity in the presence
of nonlocal interactions.

As observed in Ref.~\cite{Krasnikov:2024fdu} for the case of nonlocal
photon--$A'$ kinetic mixing, this kind of  behavior also allows
one to reconcile the stringent limits from direct dark-matter searches
with the observed relic abundance. The same occurs in our nonlocal
Stueckelberg portal scenario. 
Indeed, direct detection of the dark fermion $\chi$ proceeds through
elastic scattering $\chi + {\rm SM} \to \chi + {\rm SM}$
mediated by $t$-channel $A'$ exchange. Since the typical momentum transfer
$q^2$ in direct-detection experiments is very small, the corresponding
scattering amplitude is strongly suppressed by the nonlocal form factor
$\epsilon(q^2)$ in Eq.~\eqref{eq:NL-portal-2}, which vanishes in
the low-momentum-transfer limit. By contrast, thermal annihilation
$\chi\chi \to {\rm SM}$ in the early Universe is dominated by the $s$-channel
near-resonant or on-shell $A'$ contribution and is controlled by momentum scales
of the order of $m_{A'}^2$. Therefore, the annihilation cross section can remain
sufficiently large to reproduce the observed dark-matter relic density
while the direct-detection rate is naturally suppressed. 

Invisible and semi-invisible meson decays considered in the previous sections
are tightly constrained by fixed-target
experiments employing missing-energy and missing-momentum techniques.
For fully invisible final states, the strongest bounds come from the
{\it BABAR} search~\cite{BaBar:2009gco} and from the NA64 program
at the CERN Super Proton Synchrotron 
using electron and positron beams~\cite{NA64:2023wbi}.
In addition, projected sensitivities based on the expected $\pi^0$, $\eta$,
and $\eta'$ yields in NA64 with a negative-pion beam were estimated
in Ref.~\cite{Gninenko:2023rbf}. For the invisible decay
$\rho^0 \to \chi\bar{\chi}$, constraints derived from the $\rho$-meson
yields---computed within a Glauber-model treatment for NA64---were presented
in Ref.~\cite{Schuster:2021mlr}.  
We used these experimental constraints to analyze the numerical impact of
the nonlocality form factor~(\ref{eq:NL-portal-1}) on the meson  decay rates
derived in the previous sections and estimated the corresponding limits
on the scale $\Lambda_{\rm NL}$. 

We found that the decay rates of both vector
$V=\rho,\omega,\phi,J/\psi,\Upsilon$ and pseudoscalar $P=\pi^0,\eta,\eta'$
mesons  probe the effective  nonlocal couplings
at timelike momentum transfer, $\hat{\epsilon}(k^2=m_{A'}^2)$, implying that
the corresponding bounds become suppressed in the nonlocal scenario when
$\Lambda_{\rm NL} \gg m_M$. Consequently, limits derived from invisible 
meson decay searches rapidly weaken due to the momentum-dependent suppression
$\hat{\epsilon}(k^2)\propto (k^2/\Lambda_{\rm NL}^2)^2$.
As a result, for $\Lambda_{\rm NL}>1$TeV the Dark Photon mass $m_{A'}$ becomes
unconstrained both from meson decays and from cosmological and collider data
in Fig.~\ref{fig:boundslambda}. We already noted in Sec.~\ref{sec:setup}
that values of the nonlocality scale $\Lambda_{\rm NL}$ below the TeV range
are challenging from the viewpoint of theoretical consistency and
the construction of a sensible UV completion. 

\section{Conclusions}
\label{sec:Conclusions}

We have developed a meson-level effective description for a light vector boson
$A'_\mu$ that couples directly to SM quark and lepton currents through generic
vector and axial-vector interactions. In this class of models, the dominant
hadronic production and decay channels of $A'$ at sub- to few-GeV energies are
governed by how the underlying quark currents are embedded into low-energy QCD,
so rare meson processes provide an essential---and often leading---set
of probes.

Starting from the quark-level interactions in Eq.~(\ref{eq:Lag-SSB-1}), we
constructed an effective theory for pseudoscalar mesons supplemented by
vector-meson resonances, Eq.~(\ref{eq:Lag-eff-V-P-1}). The matching is realized
through resonance saturation: Vector quark currents generate $A'$ couplings
mediated by vector mesons and their mixing, providing a practical description
in the GeV regime. We also incorporated the anomalous WZW 
terms that govern radiative and anomalous transitions such as $P\to \gamma A'$,
$V\to P A'$, and related channels, and expressed the corresponding effective
couplings in terms of the underlying quark-level parameters. Within this unified
framework we derived the decay widths and branching ratios relevant for both
visible and invisible scenarios. For visible $A'$ decays, narrow-resonance
searches in dilepton spectra map directly onto meson transition rates, while for
invisible $A'$ decays the same meson transitions lead to missing-energy or
missing-mass signatures, enabling stringent constraints on $A'$ models.

We also explored the phenomenological viability of the nonlocal setup, assuming
that the SM--dark--sector portal is nonlocal and characterized by a nonlocality
scale $\Lambda_{\rm NL}$ controlling the effective $A'$--SM--fermion interaction.
In particular, for values of the nonlocality scale in the UV-controlled regime,
$\Lambda_{\rm NL}\gtrsim 1~\mathrm{TeV}$, the portal suppression renders existing
meson-decay, cosmological, and collider limits  insensitive to the
Dark Photon mass $m_A'$, leaving it unconstrained, at least for the benchmark
assumptions adopted in our analysis. 

Let us finally comment on the possible observability of nonlocal effects
in the present scenario. 
The scale $\Lambda_{\rm NL}$ entering our framework parametrizes specifically
the nonlocality of the portal interaction between the Standard Model and the
dark sector, rather than a universal nonlocal modification of all
Standard Model interactions. Therefore, the existing ATLAS and CMS
bounds \cite{ParticleDataGroup:2024cfk} on generic nonlocal scales at
the level of several TeV should be viewed as an important reference scale 
but not necessarily as a direct exclusion of the scenario considered here.

Note that if the effective portal nonlocality scale is extremely high,
for example, close to the Planck scale, the resulting effects become
negligibly small and the model effectively reduces to the local limit.
On the other hand, if $\Lambda_{\rm NL}$ lies in the multi-TeV range,
observable effects of nonlocality may still be accessible in rare meson
decays and missing-energy experiments.

The characteristic signature of nonlocality is the momentum dependence of
the effective $A'$ interactions encoded in the form factor $\epsilon(q^2)$.
This can lead to correlated distortions of production rates and spectra
in processes probing different momentum transfers. Our analysis demonstrates
that present and future meson-decay searches can probe a substantial part
of the corresponding parameter space.

At the same time, we note that momentum-dependent effective interactions may
also emerge in certain local UV-complete scenarios after integrating out
additional heavy degrees of freedom. Therefore, establishing experimentally
the genuinely nonlocal origin of such effects may represent a nontrivial
challenge.

In conclusion, the effective description presented here provides a systematic
and flexible framework for embedding nonlocality in quark--$A'$ interactions
into meson-level observables, and can be straightforwardly extended to refined
hadronic inputs and dedicated experimental sensitivity studies.

\begin{acknowledgments} 

This work was funded by FONDECYT (Chile) under Grants
No. 1230160 and No. 1240066, and
by ANID$-$Millen\-nium Program$-$ICN2019\_044 (Chile).
The work of A.~S.~Zh. is supported by the Foundation 
for the Advancement of Theoretical Physics and Mathematics "BASIS"
and by PIFI CAS Grant No.:2024PVB0070. 
A.~S.~Zh. also is grateful Xurong Chen (IMP CAS, China)
and Pengming Zhang (Sun Yat-Sen University, China)
for the warm hospitality, where part of work was done. 

\end{acknowledgments}

\hspace*{1cm}
  
\centerline{\bf Data Availability}

\hspace*{.25cm}

There are no publicly available research data or software supporting
this manuscript. Requests for further information or data should be
sent to the authors.

\end{document}